\documentclass[a4paper,11pt]{article}
\usepackage{amsmath}
\usepackage{jheppub}
\usepackage{amssymb}
\usepackage{graphicx,cleveref}
\usepackage[T1]{fontenc}
\usepackage{slashed}
\usepackage[utf8]{inputenc}
\usepackage{xspace}
\usepackage{color}
\usepackage{ulem}
\usepackage{array}
\usepackage{ulem,fancyvrb}
\usepackage{xcolor}

\def \abs#1{\left\vert#1\right\vert}
\allowdisplaybreaks
\date{\today}

\begin{document}

\title{A new way to test the WIMP dark matter models}

\author[1,2]{Wei Cheng,}
\author[1]{Yuan He,}
\author[1]{Jing-Wang Diao,}
\author[1]{Yu Pan,$^{\dag,}$}
\emailAdd{panyu@cqupt.edu.cn}
\author[3]{Jun Zeng$^{\dag,}$}
\emailAdd{zengj@cqu.edu.cn}
\author[4]{and Jia-Wei Zhang}
\affiliation[1]{School of Science, Chongqing University of Posts and Telecommunications, Chongqing 400065, P. R. China}
\affiliation[2]{State Key Laboratory of Theoretical Physics, Institute of Theoretical Physics, Chinese Academy of Sciences, Beijing, 100190, P.R. China}
\affiliation[3]{Department of Physics, Chongqing University, Chongqing 401331, P.R. China}
\affiliation[4]{Department of Physics, Chongqing University of Science and Technology, Chongqing 401331, P. R. China}
\affiliation[\dag]{\rm Corresponding authors:}

\abstract{In this paper, we investigate the possibility of testing
the weakly interacting massive particle (WIMP) dark matter (DM)
models by applying the simplest phenomenological model which
introduces an interaction term between dark energy (DE) and WIMP DM,
i.e., $Q = 3\gamma_{DM} H\rho_{DM}$. In general, the coupling
strength $\gamma_{DE}$ is close to $0$ as the interaction between DE
and WIMP DM is very weak, thus the effect of $\gamma_ {DE}$ on the
evolution of $Y$ associated with DM energy density can be safely
neglected. Meanwhile, our numerical calculation also indicates that
$x_f\approx20$ is associated with DM freeze-out temperature, which
is the same as the vanishing interaction scenario. As for DM relic
density, it will be magnified by $\frac{2-3\gamma_{DM}}{2}[{2\pi g_*
m_{DM}^3}/{(45 s_0 x_f^3})]^{\gamma_{DM}}$ times, which provides a
new way to test WIMP DM models. As an example, we analyze the case
in which WIMP DM is a scalar DM. (SGL+SNe+Hz) and (CMB+BAO+SNe)
cosmological observations will give
$\gamma_{DM}=0.134^{+0.17}_{-0.069}$ and
$\gamma_{DM}=-0.0008\pm0.0016$, respectively. After further
considering the constraints from DM direct detection experiment, DM
indirect detection experiment, and DM relic density, we find that
the allowed parameter space of the scalar DM model will be
completely excluded for the former cosmological observations, while
it will increase for the latter ones. Those two cosmological
observations lead to an almost paradoxical conclusion. Therefore,
one could expect more stringent constraints on the WMIP DM models,
with the accumulation of more accurate cosmological observations in
the near future.} \maketitle \preprint{}

\section{Introduction}

Dark energy (DE) and dark matter (DM), which are the main parts of
the energy content of the universe, have been firmly established by
numerous astronomical and cosmological
observations~\cite{Blumenthal:1984bp,Davis:1985rj,Clowe:2006eq,Riess:1998cb,Perlmutter:1998np,Cao2013a,Cao2014,Cao2017a,Cao2018a}.
More specifically, the cosmic microwave background (CMB)
anisotropies have verified the abundance of DM and DE with
remarkable precision at $(26.0 \pm
0.5)\%$~\cite{Bennett:2012zja,Ade:2015xua} and $(68.89 \pm
0.56)\%$~\cite{Aghanim:2018eyx} respectively. The usual strategy of
particle physicists to deal with the DM is to extend the standard
model (SM) of particle physics by adding a new particle that
interacts with the SM particles. Based on this interaction,
physicists have conducted indirect
detection~\cite{Heinke:2014hoa,Bringmann:2012vr,Weniger:2012tx,Macias:2019omb,FermiLAT:2011ab},
direct
detection~\cite{Aprile:2012zx,Akerib:2013tjd,Akerib:2016vxi,Tan:2016zwf},
and collider
detection~\cite{Fujii:2015jha,Gomez-Ceballos:2013zzn,dEnterria:2016sca,CEPC-SPPCStudyGroup:2015csa}
for testing DM models.

There is plentiful literature devoted to the exploration of the
interaction between DE and DM to relieve the coincidence
problem~\cite{Zimdahl:2001ar,Feng:2008fx,Wang:2006qw,Wang:2005jx,Cui:2010dr,Abdalla:2007hc,Jamil:2008rc,
Bertolami:2007zm,Szydlowski:2005ph,Chen:2010dk,Cao:2010fb,Zhang:2009ay,Zhang:2010icb,Kumar:2017bpv,Yang:2018euj,DiValentino:2019ffd,DiValentino:2019jae}.
More specifically, in Ref.~\cite{Cao2013b,Pan:2012qt,Pan2016}, the
interaction type between DM and DE is set as $ Q = 3H \gamma_{m}
\rho_m$. After considering the limits from the combination of
cosmological data (GRBs+SNe+BAO+CMB), the coupling strength of DM
and DE can be obtained, i.e., $\gamma_{m}= -0.0047 \pm 0.0046$ in $1
\sigma$ errors, which indicates a slight energy transfer from DM to
DE. In Ref.~\cite{Cao:2014moa}, the interaction between DE and DM
(baryonic substances) is set as $3H \gamma_{c} \rho_c$ ($3H
\gamma_{b} \rho_b$). Taking the bounds from the combination of
cosmological data (SNe+BAO+Planck+Hz), $\gamma_{c} = 0.0078 \pm
0.0045$ and $\gamma_{b}= 0.0030^{+0.0255}_{-0.0125}$ in $1 \sigma$
errors can be obtained. While using the bounds from the combinations
of cosmological data (SNe+BAO+WMAP9+Hz), one can get $\gamma_{c} =
0.0068^{+0.0043}_{- 0.0042}$ and $\gamma_{b} =
0.0025^{+0.0235}_{-0.0115}$ in $1 \sigma$ errors. If one sets
$\gamma_{b}=0$, then $\gamma_{c}=0.0095\pm 0.0031 $ in $1 \sigma$
errors within the bounds from the combinations of cosmological data
(SNe+BAO+Planck+Hz), and $\gamma_{c}=0.0081\pm0.0031 $ in $1 \sigma$
errors within the bounds from the combinations of cosmological data
(SNe+BAO+WMAP9+Hz). All predictions of Ref.~\cite{Cao:2014moa}
indicate $\gamma_{c}>0$, which means the energy of DE is slightly
transferring to that of DM. However, These two references imply that
the cosmological data cannot completely exclude the interaction
between DE and DM, and different cosmological data may lead to
different prediction about the energy transference between DE and
DM.

Motivated by the method of DM Model detection which is based on the
interaction between the SM particle and new
particles~\cite{Cao:2012fz,Cao:2007rm,Gao:2015irw,Terazawa:2015bsa,Liu:2017rgs,Yepes:2018zkk,Cao2021},
we will attempt to develop an alternative DM model verification
method based on the new interaction between DM and DE. In the new
model, we will restrict ourselves to the Weakly Interacting Massive
Particle (WIMP) DM. Specifically, we studied the simplest type of
interaction between DE and WIMP DM, namely $Q = 3\gamma_{DM}
H\rho_{M}$. We deduced the WIMP DM relic density that can be used to
examine the WIMP DM models with the help of DM detection
experiments.

As an example, we will describe the WIMP DM as a real scalar
particle that is the simplest Higgs-portal DM model and has had been
comprehensively
revisited~\cite{Silveira:1985rk,McDonald:1993ex,Burgess:2000yq,Barger:2007im,Gonderinger:2009jp,Han:2016gyy,Wu:2016mbe,Boehm:2020wbt}.
After considering the WIMP DM relic density, the parameter space of
the scalar model will increase with $\gamma_{DM}<0$, and the smaller
$\gamma_{DM}$, the larger the increased parameter space of the
model; while the parameter space will shrink with $\gamma_{DM}>0$,
and the larger $\gamma_{DM}$, the larger excluded parameter space of
this model.

The remaining parts of this paper are organized as follows.
In Sec.~\ref{sec:II}, we construct the DE and WIMP DM models with an interaction term $Q = 3\gamma_{DM} H\rho_{M}$.
In Sec.~\ref{sec:III}, we show three observational data in Cosmology and further discuss the coupling strength $\gamma_{DM}$.
In Sec.~\ref{sec:IV}, the WIMP DM relic density in the new models is calculated.
As an example, a scalar DM as the WIMP DM is investigated in detail in Sec.~\ref{sec:V}.
Finally, we will briefly summarize in Sec.~\ref{sec:VI}.

\section{Dark energy and WIMP dark matter interaction model}\label{sec:II}

The energy evolution of DE and matter can be described as follows~\cite{Cao:2014moa,Pan:2015quf}:
\begin{eqnarray}{\label{EQ:DEM}}
\dot{\rho}_{DE} + 3H(\rho_{DE} + P_{DE})&=& - Q, \nonumber \\
\dot{\rho}_{M} + 3H\rho_{M}&=&  Q.
\end{eqnarray}
where $\rho_{DE}$ and $\rho_{M}$ are the DE and matter energy respectively. The interaction term $Q$ describes the interchange of energy with each other, but there is no interaction between them when $Q=0$. Note that those two energies meet the total energy conservation equation: $\dot{\rho}_{tot} + 3H(\rho_{tot} + P_{tot}) = 0$.

We will further divide the matter energy into DM and the usual standard model matter, i.e. $\rho_{M}=\rho_{DM}+\rho_{SM}$. If the DM is the WIMP, then the evolution of $\rho_{DM}$ can be written as follows:
\begin{eqnarray}
\dot{\rho}_{DM} + 3H\rho_{DM}&=&  Q_{DM} + \frac{\langle \sigma v \rangle}{m_{DM}}(\rho_{DM,eq}^2-\rho_{DM}^2), \label{EQ:DM}
\end{eqnarray}
where $\langle \sigma v \rangle$ is the WIMP DM thermal average cross section, and $\rho_{DM,eq}$ represents the WIMP DM energy of thermal equilibrium state.
Note that Eq.~(\ref{EQ:DM}), which is different from other papers, is the most important innovative part of our paper, as both of the WIMP DM annihilation term and the interaction term of DM and DE are considered simultaneously. Usually, cosmologists may not be concerned with the annihilation term $\frac{\langle \sigma v \rangle}{m_{DM}}(\rho_{DM,eq}^2-\rho_{DM}^2)$, while particle physicists do not consider the interaction term $Q_{DM}$.

Following the common practice in extensive literature, we will consider the usual scenario interaction term $Q = 3\gamma_{DM} H\rho_{M}$ and assume $Q_{DM} = 3\gamma_{DM} H\rho_{DM}$. When $\gamma_{DM}<0$, the energy is transferred from WIMP DM to DE, while for $\gamma_{DM}>0$, the energy is transferred from DE to WIMP DM, which will affect the feasible parameters of the WIMP DM model. Specifically, the feasible parameters of the WIMP DM model will be shrunken for the case $\gamma_{DM}>0$, while it will be increased for the case $\gamma_{DM}<0$. Furthermore, the standard $\Lambda$CDM model without interaction between DE and matter is characterized by $Q  = 0$, while $Q\neq 0$ represents the non-standard cosmology with interaction between DE and matter.

With the above Eq.~(\ref{EQ:DM}) at hand, we can establish a relation between the coupling strength $\gamma_{DM}$ and the thermal average cross section of WIMP DM through its relic density. As the cosmology observations make a constraint for the coupling strength $\gamma_{DM}$ through Eq~.(\ref{EQ:DEM}), we can also establish a relation between the cosmology observations and the thermal average cross section of WIMP DM. We will discuss it in detail in the following parts.

\section{Observational constraints on $\gamma_{DM}$}\label{sec:III}

In this section, we will introduce three types of observational data
to place constraints on the $\gamma_{DM}$ interaction dark energy
model parameters, i.e., 130 updated galaxy-scale strong
gravitational lensing sample (SGL)~\citep{Cao2015,chenyun:2019qso}
with redshift from $0.197$ to $2.834$, Hubble parameter data
(Hz)~\citep{Wei:2017Hz}, and Pantheon 1048 Ia supernovae sample
(SNe) discovered by the Pan-STARRSI Medium Deep
Survey~\citep{Scolnic:2018Pantheon}.

\subsection{Observational data in cosmology}

In the electromagnetic and gravitational wave domain
\cite{Cao2015,chenyun:2019qso,Bian2021,Ola2021}, strong
gravitational lensing has been widely used in precision cosmology,
concerning accurate determination of cosmological parameters
\cite{Chen2015,Qi2018a,Liu2019,Liu2020,Qi2021} and cosmic opacity at
higher redshifts \cite{Ma2019a}, constraints on the velocity
dispersion function \cite{Cao2012,Ma2019b,Geng2021} and dark matter
distribution in early-type galaxies \cite{Cao2016}, direst tests of
Parametrized Post-Newtonian gravity \cite{Cao2017b} and the validity
of the FLRW metric \cite{Cao2019,Qi2019a}, as well as precise
measurements of the speed of light \cite{Cao2018b,Cao2020,Liu2021a}
with galaxy-scale strong lensing systems.

Considering the influence of some unknown system errors on the SGL
data, Chen et al. compiled $161$ galaxy-scale strong gravitational
lensing sample system which include the gravitational lensing and
stellar velocity dispersion measurements. They selected those
samples from early-type galaxies with $E/S0$ morphologies with
strict criteria to satisfy the assumption of spherical symmetry on
the lens mass model. In their model, they discuss the slope of the
total mass density profile $\gamma$ considering three
parameterizations ($\gamma=\gamma_0$,
$\gamma=\gamma_0+\gamma_z\times z_l$ and
$\gamma=\gamma_0+\gamma_z\times z_l+\gamma_s\times
log\tilde{\sum}$.), the luminosity density slope $\delta$ with an
observable parameter for each lens and the orbit anisotropy
parameter $\beta$ treated as a nuisance parameter and marginalized
over with a Gaussian prior $\beta=0.18\pm0.13$
\citep{chenyun:2019qso}. Moreover, because of the high-resolution
HST imaging data needed, they finally chose $130$ galaxy-scale
strong gravitational lensing data from $161$ samples, which are
separated from the Sloan Lens ACS(SLACS) and others surveys. The
specific selection methods can be found in the literature
\citep{chenyun:2019qso}.

When considering the effect of aperture size on the velocity
dispersion of the lens galaxy, a more appropriate choice for the
radius is $R_{\text {eff }}/2$ with $R_{\text {eff}}$ being the
half-light radius of the lens galaxy, because $R_{\text {eff }}$
matches the Einstein radius well~\citep{Auger:2010sigma}. Choose
$R_{\text {eff}}/2$ as the radius to get the velocity dispersion
$\sigma_{e2}$, according to the regulations, the observational
expression of velocity dispersion is as follows \cite{Cao2015}
\begin{eqnarray}\label{Quasarobs}
\sigma_{\mathrm{obs}} \equiv \sigma_{e2} =\sigma_{ap}\left[\theta_{\mathrm{eff}} /\left(2 \theta_{a p}\right)\right]^{\eta},
\end{eqnarray}
where $\sigma_{ap}$ is velocity dispersion, $\theta_{\text {eff }}=R_{\text {eff }} / D_{l}$, the correction factor $\eta=-0.066\pm0.035$~\citep{Cappellari:2006eta}, and $\theta_{\mathrm{ap}} \approx 1.025 \times \sqrt{\left(\theta_{x} \theta_{y} / \pi\right)}$ with $\theta_{x}$ and $\theta_{y}$ being the angular sizes of width and length of the rectangular aperture respectively~\citep{Jorgensen:1005sita}.

The total error of the actual velocity dispersion is
\begin{eqnarray}\label{Quasarerr}
\left(\Delta \sigma_{\mathrm{tot}}\right)^{2}=\left(\Delta \sigma_{\mathrm{stat}}\right)^{2}+\left(\Delta \sigma_{\mathrm{sys}}\right)^{2}+\left(\Delta \sigma_{\mathrm{AC}}\right)^{2},
\end{eqnarray}
where $\Delta \sigma_{\mathrm{stal}}$, propagated from the measurement error of $\sigma_{\mathrm{ap}}$, is the statistical error, $\Delta \sigma_{\mathrm{gys}}$ is systematic error, and the error $\Delta \sigma_{\mathrm{AC}}$ is propagated from the uncertainty of $\eta$ due to the aperture correction.

The theoretical expression of the velocity dispersion can be written
as \cite{Cao2016}
\begin{eqnarray}\label{Quasarth}
\sigma_{th}=\sqrt{\frac{c^{2}}{2 \sqrt{\pi}} \frac{D_{s}}{D_{l s}} \theta_{E} \frac{3-\delta}{(\xi-2 \beta)(3-\xi)} F(\gamma, \delta, \beta)\left(\frac{\theta_{\mathrm{eff}}}{2 \theta_{\mathrm{E}}}\right)^{(2-\gamma)}},
\end{eqnarray}
where $D_{ls}$ is the angular diameter distance between lens and source, $D_s$ is that between observer and source, which are dependent on the cosmological model. The cosmological model enters into the theoretical observable not through a distance measure directly, but rather through a distance ratio:$\frac{D_{s}\left(z_{s} ; \mathbf{p}, H_{0}\right)}{D_{l s}\left(z_{l}, z_{s} ; \mathbf{p}, H_{0}\right)}$. The expressions for both $D_{s}\left(z_{s} ; \mathbf{p}, H_{0}\right)$ and $D_{l s}\left(z_{l}, z_{s} ; \mathbf{p}, H_{0}\right)$ are as follows:
\begin{eqnarray}\label{Ds}
D_{s}\left(z_{s} ; \mathbf{p}, H_{0}\right)=\frac{c}{H_{0}\left(1+z_{s}\right)} \int_{0}^{z_{s}} \frac{d z}{E(z ; \mathbf{p})},
\end{eqnarray}
\begin{eqnarray}\label{Dls}
D_{l s}\left(z_{l}, z_{s} ; \mathbf{p}, H_{0}\right)=\frac{c}{H_{0}\left(1+z_{s}\right)} \int_{z_{l}}^{z_{s}} \frac{d z}{E(z ; \mathbf{p})},
\end{eqnarray}
where $E(z ; \mathbf{p})$ is the cosmological background of the
$\gamma_{DM}$ model that has the following form
\begin{eqnarray}\label{Ez2}
E^{2}(z;\mathbf{p})=\frac{w_{x} \Omega_{m}}{\gamma_{DM}+w_{x}}(1+z)^{3\left(1-\gamma_{DM}\right)}+\left(1-\frac{w_{x} \Omega_{m}}{\gamma_{DM}+w_{x}}\right)(1+z)^{3\left(1+w_{x}\right)}
\end{eqnarray}
where $\mathbf{p}$ is the model parameters, i.e., $\mathbf{p}=(\Omega_m,\omega_x,\gamma_{DM})$.

The function $F(\gamma, \delta, \beta)$ in Eq.(\ref{Quasarth}) can
be expressed as \cite{Cao2015}
\begin{eqnarray}\label{QuasarthF}
F(\gamma, \delta, \beta)=\left[\frac{\Gamma[(\xi-1) / 2]}{\Gamma(\xi / 2)}-\beta \frac{\Gamma[(\xi+1) / 2]}{\Gamma[(\xi+2) / 2]}\right] \frac{\Gamma(\gamma / 2) \Gamma(\delta / 2)}{\Gamma[(\gamma-1) / 2] \Gamma[(\delta-1) / 2]}.
\end{eqnarray}
where $\xi = \gamma + \delta -2$. In this paper, we consider the dependence of $\gamma$ on lens redshift $z_l$ and the dependence of redshift on surface mass density (i.e., P3 in the literature\citep{chenyun:2019qso}). The expression of $\gamma$ is as follows:
\begin{eqnarray}\label{gamma}
\gamma=\gamma_{0}+\gamma_{z} \times z_{l}+\gamma_{s} \times \log \tilde{\Sigma},
\end{eqnarray}
where $\tilde{\Sigma}$ is the surface mass density,
\begin{eqnarray}\label{E}
\tilde{\Sigma}=\frac{\left(\sigma_{obs} / 100~\mathrm{km} \mathrm{s}^{-1}\right)^{2}}{R_{\mathrm{eff}} / 10 h^{-1}~\mathrm{kpc}}
\end{eqnarray}
$h$ is from the Hubble constant $H_0=100h~\rm{km}^{-1}\rm{Mpc}^{-1}$.

In addition, we also consider the latest Pantheon supernova sample~\citep{Scolnic:2018Pantheon}
\begin{eqnarray}\label{mu}
\mu=M + 5 \log \frac{d_{L}}{\mathrm{pc}}+25,
\end{eqnarray}
where
\begin{eqnarray}\label{DL}
d_{L}=\frac{(1+z)}{H_{0}} \int_{0}^{z} \frac{d z^{\prime}}{E\left(z^{\prime}\right)}.
\end{eqnarray}

Finally, the 31 Hubble parameter (Hz) samples from the differential
age method are also used~\citep{Wei:2017Hz}, focusing on
model-independent reconstruction of cosmological distances
\citep{Wu2020,Zheng2021a} and the spatial curvature of the Universe
\cite{Liuyu2020,Zheng2021b}, as well as statistical analysis of
parameterized $Om(z)$ diagnostics in light of recent observations
\cite{Zheng2016,Qi2018b}.

\subsection{Constraint on $\gamma_{DM}$}

The parameters of the $\gamma_{DM}$ model are constrained by using (SGL+SNe+Hz) sample combination through Markov chain Monte Carlo(MCMC) method~\citep{Lewis02}, and fitted by the minimum likelihood method of $\chi^2$. The final $\chi_{All}^2$ is given by the following function:
\begin{eqnarray}\label{allchi2}
\chi_{All}^{2}=\chi_{SGL}^{2}+\chi_{SNe}^{2}+\chi_{Hz}^{2}.
\end{eqnarray}
where $\chi_{SGL}^{2}$, $\chi_{SNe}^{2}$, and $\chi_{Hz}^{2}$ can be expressed as follows:
\begin{eqnarray}\label{Quasarchi2}
\chi_{SGL}^{2}=\sum_{i=1}^{130}\left(\frac{\sigma_{\mathrm{th}}-\sigma_{\mathrm{obs}}}{\Delta \sigma_{\mathrm{tot}}}\right)^{2},
\end{eqnarray}
\begin{eqnarray}\label{Pantheonchi2}
\chi_{SNe}^{2}=\sum_{i=1}^{1048}\left(\frac{\mu_{th}-\mu_{obs}}{\sigma_{\mu}}\right)^{2},
\end{eqnarray}
\begin{eqnarray}\label{Hzchi2}
\chi_{Hz}^{2}=\sum_{i=1}^{31}\left(\frac{Hz_{th}-Hz_{obs}}{\sigma_{Hz}}\right)^{2}.
\end{eqnarray}

\begin{figure}[t]
\begin{centering}
\includegraphics[width=0.85\textwidth,height=0.9\textwidth]{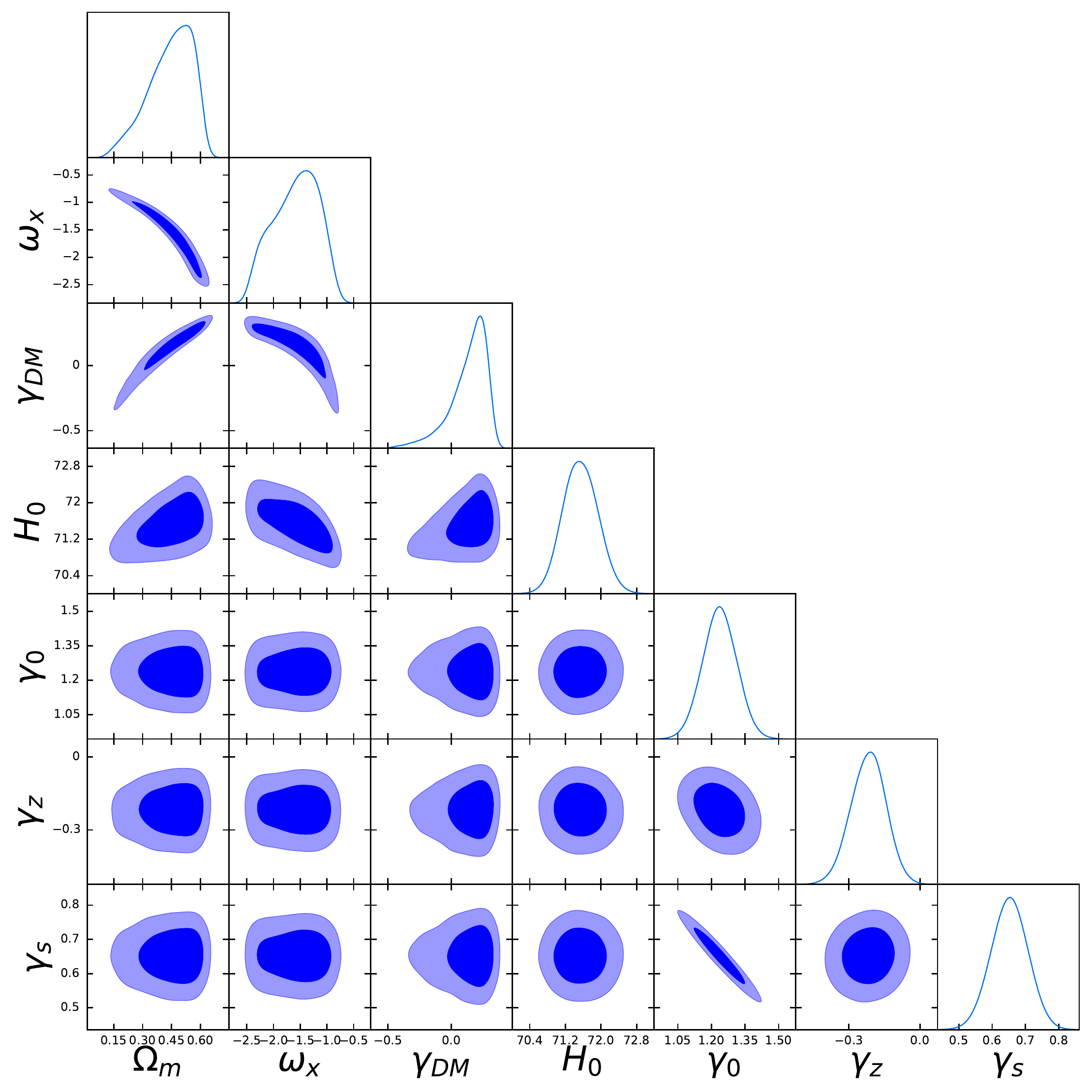}
\caption{The 2D contour line are obtained by using SGL+SNe+Hz.}
\label{Fig:Alldata}
\end{centering}
\end{figure}

The best values of each parameter ($\Omega_m$, $\omega_x$, $\gamma_{DM}$, $H_0$, $\gamma_0$, $\gamma_z$, $\gamma_s$) and their $2\sigma$ error results are shown in Table.\ref{gammaIDE}. The 2D contour line is shown in Fig.\ref{Fig:Alldata}.

\begin{table}
\begin{center}
\begin{tabular}{ccc}
\hline
Parameters         &~~1$\sigma$~~                 &~~2$\sigma$~~~~\\
\hline
$\Omega_m$         &$0.445_{-0.076}^{+0.15}$      &$0.445_{-0.23}^{+0.19}$\\
$\omega_x$         &$-1.59_{-0.35}^{+0.54}$       &$-1.59_{-0.82}^{+0.71}$\\
$\gamma_{DM}$      &$0.134_{-0.069}^{+0.17}$      &$0.134_{-0.32}^{+0.23}$\\
$H_0$              &$71.54\pm0.39$                &$71.54_{-0.74}^{+0.78}$\\
$\gamma_0$         &$1.237\pm0.067$               &$1.237_{-0.15}^{+0.15}$\\
$\gamma_z$         &$-0.218\pm0.074$              &$-0.218_{-0.15}^{+0.14}$\\
$\gamma_s$         &$0.652\pm0.055$               &$0.652_{-0.11}^{+0.11}$\\
\hline
\end{tabular}
\caption{The best values ($\Omega_m$, $\omega_x$, $\gamma_{DM}$, $H_0$, $\gamma_0$, $\gamma_z$, $\gamma_s$) and their $2\sigma$ errors of the $\gamma_{DM}$ model  parameters are obtained by using SGL+SNe+Hz observational data combination.}
\label{gammaIDE}
\end{center}
\end{table}

We can see that the interaction factor $\gamma_{DM}$
($\gamma_{DM}=0.134_{-0.069}^{+0.17}$) is not zero within the
$1\sigma$ error range and the best value is also $\gamma_{DM}>0$
from Table~\ref{gammaIDE}, which indicate there is a trend of
conversion from DE to DM.  The coincidence problem may be alleviated
slightly at the range of $1\sigma$ error. In addition, the best
value of $H_0$ is $H_0=71.54~\rm{kms}^{-1}\rm{Mpc}^{-1}$, which is
larger than $H_0=67.4~\rm{kms}^{-1}\rm{Mpc}^{-1}$ given by
Plank2018~\citep{Aghanim:2018eyx} and smaller than
$H_0=74.03~\rm{kms}^{-1}\rm{Mpc}^{-1}$ given by
Pantheon~\citep{Riess:2019H0SNeCMB}, which alleviates the conflict
problem of $H_0$ to some extent. It shows that the $\gamma_{DM}$
model has the ability to alleviate the tension problem of Hubble
constant $H_0$ which is consistent with the results in
reference~\citep{zhengxg17}. We also note that the coupling
parameter $\gamma_{DM}$ is correlated with all other model
parameters $\Omega_m$, $\omega_X$ and $H_0$, and the cosmological
constant model ($\Lambda$CDM) ($\gamma_{DM}=0$, $\omega_X=0$) is
contained in the $2\sigma$ confidence region. The matter density
parameter $\Omega_m$ is also consistent with the result from other
observational data in the $2\sigma$ confidence
region~\citep{Aghanim:2018eyx, Riess:2019H0SNeCMB} although the best
value is slightly larger.

For the large $\gamma_{DM}$ obtained from the (SGL+SNe+Hz) data analysis, we analyzed the SGL data separately, which leads to $\gamma_{DM}=0.15^{+0.35}_{-0.10}(1\sigma)^{+0.36}_{-0.51}(2\sigma)$.
Compared with $\gamma_{DM}=0.134^{+0.17}_{-0.069}(1\sigma)^{+0.23}_{-0.32}(2\sigma)$ obtained by (SGL+SNe+Hz) data analysis,
one can conclude that SGL is the main reason for the large $\gamma_{DM}$.

\begin{figure*}[h]
\begin{centering}
\includegraphics[width=0.75 \textwidth]{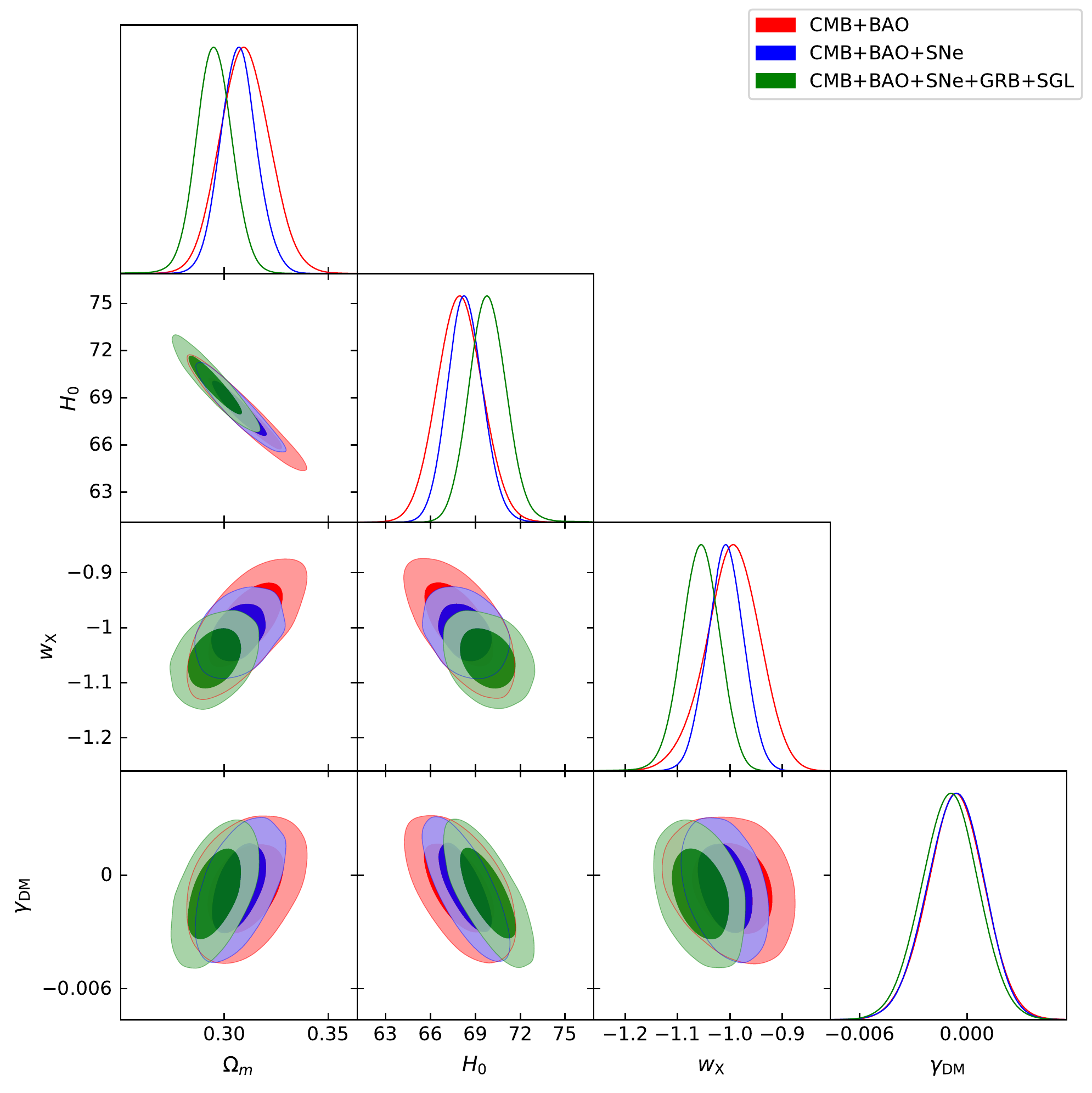}
\caption{The 2D contour line are obtained by using (CMB+BAO), (CMB+BAO+SNe) and (CMB+BAO+SNe+GRB+SGL).}
\label{Fig:all}
\end{centering}
\end{figure*}

For comparison with the standard "BAO+SNe+CMB", we also consider other combinations of these data for comparison (CMB+BAO, CMB+BAO+SNe, CMB+BAO+SNe+GRB, CMB+BAO+SNe+GRB+SGL). The total results are shown in Table~\ref{tab:my_label} and in Fig.~\ref{Fig:all}.
In which, the CMB measurement from Planck TT, TE, EE+lowE data are released in 2018~\cite{Aghanim20}, and the BAO data set used in our analysis includes five lower-redshift BAO measurements from galaxy surveys at the redshifts $z=0.106\sim0.61$ regions and one higher-redshift BAO measurements from Ly$\alpha$ forest (Ly$\alpha$) data at $z=2.35$. For the lower-redshift BAO observations, we turn to the latest measurements of the acoustic-scale distance ratio from 6dFGS at $z=0.106$~\cite{Beutler11}, the SDSS Data Release 7 Main Galaxy sample at $z=0.15$~\cite{Ross15}, three data obtained by the BOSS DR12 galaxies ($z=0.38,0.51,0.61$)~\cite{Alam17}, and the higher-redshift BAO measurement is quasar-ly$\alpha$ cross in combination with ly$\alpha$ auto combination at $z=2.35$~\cite{Michael19}.

Firstly, from the two sets of standard data (CMB+BAO) and (CMB+BAO+SNe) in Table~\ref{tab:my_label}, one can find that the model parameters ($\Omega_m$, $w_{\rm{X}}$, $\gamma_{\rm{DM}}$, $H_0$ ) are confined tightly.
Secondly, according to the Table.~\ref{tab:my_label}, the GRB data had little effect on the results, which is consistent with our previous Ref.~\cite{Pan:2012qt}. Specifically, the best value of $\gamma_{\rm{DM}}$, $\Omega_m$ and $H_0$ change slightly and the $\gamma_{\rm{DM}}$ results constrained by the (CMB+BAO+SNe) and (CMB+BAO+SNe+GRB) are almost the same at $1\sigma$ error region. Thirdly, we combine the SGL data with (CMB+BAO+SNe+GRB) data to confine the value of $\gamma_{DM}$, and the results are also listed in Table~\ref{tab:my_label} and Fig.~\ref{Fig:all}.
Although the analysis of SGL data alone leads $\gamma_{DM}$ to be larger, the value of $\gamma_{DM}$ from the analysis of (CMB+BAO) or (CMB+BAO+SNe)
is close to that of the combined analysis of SGL and (GRBs+SNe+BAO+CMB), which indicates that the data (CMB+BAO) is the strongest and standard confine.
In addition, the best value $\gamma_{DM}$ and the $1\sigma$ error region from the analysis of SGL data are excluded by the standard "BAO+SNe+CMB",
but the results are consistent with each other within $1\sigma$ region when they are constrained by combined data (CMB+BAO+SNe+GRB+SGL).
The tension of the best value $\gamma_{DM}$ obtained from SGL data and the standard "BAO+SNe+CMB" may be attributed to SGL data with
lower redshift compared with CMB data with high redshift.
Although the best values of $\gamma_{DM}$ from SGL are excluded by the standard "BAO+SNe+CMB",
$\gamma_{DM}=0$ is within the $2\sigma$ confidence region, which is consistent with the conclusion of the standard "BAO+SNe+CMB" analysis.

\begin{table}[]
    \centering
    \resizebox{\textwidth}{19mm}{
    \begin{tabular}{c|cccc}
\hline
     Parameters   &  CMB+BAO&   CMB+BAO+SNe  & CMB+BAO+SNe+GRB & CMB+BAO+SNe+GRB+SGL\\
     \hline
         $\Omega_m$          & $0.310\pm 0.012  $         & $0.3065\pm 0.0086 $ &$0.3065\pm 0.0086$  &  $0.2952\pm 0.0084 $      \\
          $w_{\rm{X}} $      &$-0.993^{+0.056}_{-0.049} $ &  $-1.008\pm 0.035 $ & $-1.007\pm 0.035  $  & $-1.058\pm 0.035  $       \\
          $\gamma_{\rm{DM}}$ & $-0.0007\pm 0.0016  $      & $-0.0008\pm 0.0016$ &$-0.0008\pm 0.0016$ &$-0.0009\pm 0.0016 $        \\
          $H_0$              & $68.0\pm 1.5 $             & $68.4\pm 1.2      $ &$68.4\pm 1.2$  & $69.8\pm 1.2   $       \\
          \hline
          $\gamma_{0}$ & -     &  - &  -    & $2.75\pm 0.13              $       \\
          $\gamma_{z}$ & -     &  - &  -    &$-0.412^{+0.10}_{-0.092}   $        \\
          $\gamma_{s}$ & -     &  - &  -    & $0.256\pm 0.054            $        \\
        \hline
    \end{tabular}}
    \caption{The best values and their 1$\sigma$ errors of the $\gamma_{\rm{DM}}$ model parameters are obtained by using (CMB+BAO), (CMB+BAO+SNe), (CMB+BAO+SNe+GRB) and (CMB+BAO+SNe+GRB+SGL) observation combinations.}
    \label{tab:my_label}
\end{table}

\section{WIMP dark matter relic density}\label{sec:IV}

The evolution of DM can be described by:
\begin{eqnarray}\label{Q1rhoDM}
\dot{\rho}_{DM} + 3H\rho_{DM}&=&  3\gamma_{DM} H\rho_{DM} + \frac{\langle \sigma v \rangle}{m_{DM}}(\rho_{DM,eq}^2-\rho_{DM}^2),
\end{eqnarray}
where $\rho_{DM}=m_{DM}n_{DM}$ with $n_{DM}$ being the DM number density. The above evolution of DM equation contains the core information of DM, such as the freeze-out parameter $x_f=m_{DM}/T$ with $T$ being the DM freeze-out temperature and the DM relic density. In the following, we will go through the derivation of these quantities in detail in this new model~\cite{berkeley}.

By making a substitution $x=m_{DM}/T$ and $Y=n_{DM}S^{\gamma_{DM}-1}$ with $S$ being the entropy of the universe, the Eq.~\eqref{Q1rhoDM} with the help of relation $dS(T)/dt+3S(T)H=0$ and $\left\langle\sigma v\right\rangle = \left\langle\sigma v\right\rangle_0 x^{-n}$ can be achieved:
\begin{eqnarray}\label{Q1rhoDMY}
\frac{dY}{dx} = \frac{\langle \sigma v \rangle_0 x^{3\gamma_{DM}-n-2}}{H(m)S(m)^{\gamma_{DM}-1}}(Y^2_{eq}- Y^2),
\end{eqnarray}
where $H(m)= H x^2$ and $S(m)=S(T) x^3$.

For the further study, we set $y= \frac{\langle \sigma v  \rangle_0 n_{DM} S^{\gamma_{DM}-1}}{H(m)S(m)^{\gamma_{DM}-1}} $, thus $y_{eq}= \frac{\langle \sigma v \rangle_0 n_{DM,eq} S^{\gamma_{DM}-1}}{H(m)S(m)^{\gamma_{DM}-1}} $ with $n_{DM,eq} = g(\frac{mT}{2\pi})^{3/2}\exp{(-m/T)}$. We can further transform Eq.~\eqref{Q1rhoDM} as:
\begin{eqnarray}
\frac{dy}{dx} = \frac{1}{x^{3-3\gamma_{DM}}}(y^2_{eq}- y^2).
\end{eqnarray}

\begin{figure}
\begin{centering}
\includegraphics[width=0.5\textwidth]{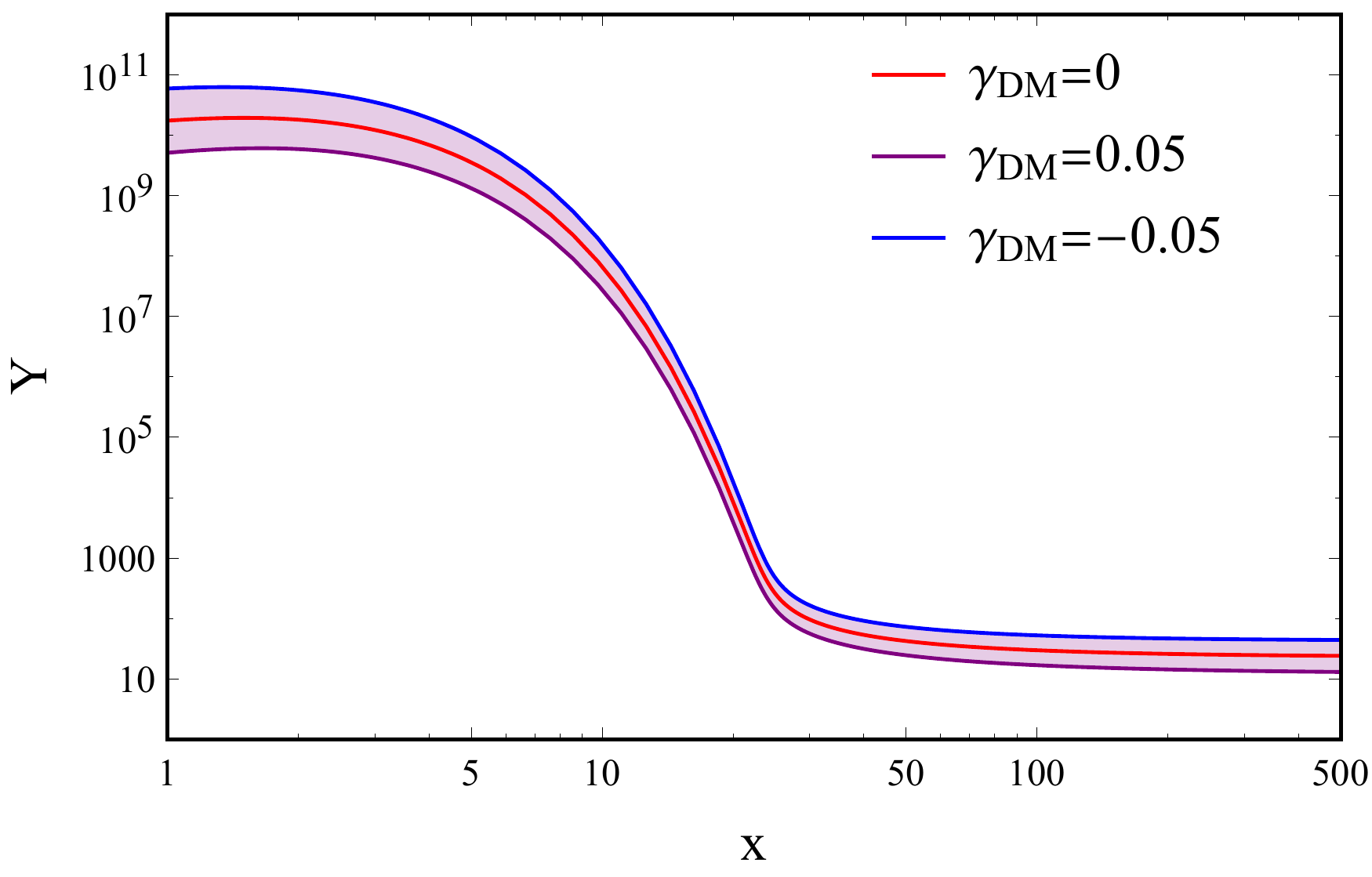}
 \caption{The plot of $Y$ by $x$ for several values of $\gamma_{DM}$.}
 \label{fig:YX}
\end{centering}
\end{figure}

The interaction strength $\gamma_{DM}$ between DE and DM, though generally small, will be set to $\gamma_{DM}<|0.05|$ to test the $x_f$. Taking the usual inputs\footnote{DM mass is $m_{DM} = 1000~\rm{GeV}$, reduced Planck mass is $M_{Pl} = (8\pi G)^{-1/2} = 2.44\times10^{18}$ $\rm{GeV}$, the number of degrees of freedom for the field is $g_* = 100$, and the cross section is $ \langle \sigma v\rangle_0= 10^{-10}$ $\rm{GeV}^{-2}$.}, we plot $Y$ as a function of $x$ in Fig.~\ref{fig:YX}. We find that $x_f$ hardly changes in $\gamma_{DM}<|0.05|$, i.e., $x_f\approx20$.

As the WIMP DM relic density is given by:
\begin{eqnarray}
\Omega = \frac{\rho_{DM_0}}{\rho_{crit}}
\end{eqnarray}
where the critical energy density $\rho_{crit}=\frac{3H_0}{8\pi G_N}$ and the energy density of DM today can be expressed as follows:
\begin{eqnarray}
\rho_{DM_0}=m_{DM}Y_{\infty}s_0^{1-\gamma_{DM}},
\end{eqnarray}
where $Y_{\infty}= \frac{H(m)S(m)^{\gamma_{DM}-1}}{\langle \sigma v \rangle_0} (n+1-3\gamma_{DM})x_f^{n+1-3\gamma_{DM}}$, which can be obtained by solving Eq.~\eqref{Q1rhoDMY}, and the entropy of the universe today is $S_0=2890~\rm{cm}^{-3}$.

Finally, the WIMP DM relic density can be obtained with $n=1$ as follows:
\begin{eqnarray}\label{EQ:Q1DMRD}
\Omega h^2 = 0.169\times x_f\sqrt{\frac{100}{g_*}}\frac{10^{-10} {GeV}^{-2}}{\langle \sigma v\rangle}\frac{2-3\gamma_{DM}}{2}\bigg(\frac{2\pi g_* m_{DM}^3}{45 S_0x_f^3}\bigg)^{\gamma_{DM}}.
\end{eqnarray}

\section{Example -- Scalar dark matter as the WIMP dark matter}\label{sec:V}

\subsection{Scalar dark matter model}
Let us firstly discuss the Higgs potential with a real scalar DM, which can be written as\cite{Wu:2016mbe}:
\begin{eqnarray}\label{V:HS}
V{(\Phi,\mathcal{S})} &=& - \mu_h^2 \abs{\Phi}^2 + \lambda_h \abs{\Phi}^4 + \frac{1}{2}\mu_\mathcal{S}^2 {\mathcal{S}}^2 + \frac{1}{2} \lambda_\mathcal{S} {\mathcal{S}}^4 + \frac{1}{2} \lambda_{h\mathcal{S}}\abs{\Phi}^2{\mathcal{S}}^2,
\end{eqnarray}
where $\Phi$ and $\mathcal{S}$ are the SM Higgs and the real scalar DM, respectively. The interaction between DM sector and SM Higgs is described by the last term, and the coupling constant $\lambda_{h\mathcal{S}}$ reflect the strength of this interaction.
The cubic term of $\mathcal{S}$ is eliminated due to the even $Z_2$ parity symmetry of the real scalar DM.

After the electroweak symmetry spontaneously breaks, the dark sector of Eq.~(\ref{V:HS}) can be written as,
\begin{eqnarray}{\label{Lag2}}
V(h,\mathcal{S})&=&  \frac{m_{\mathcal{S}}^{2}}{2} \mathcal{S}^{2} + \frac{\lambda_\mathcal{S}}{2}\mathcal{S}^{4} + \frac{\lambda_{h\mathcal{S}} \upsilon}{2}\mathcal{S}^{2}h + \frac{\lambda_{h\mathcal{S}}}{4}\mathcal{S}^{2}h^{2},
\end{eqnarray}
where the square of DM mass is $m^{2}_{\mathcal{S}}=\mu_\mathcal{S}^{2}+\lambda_{h\mathcal{S}} \upsilon^{2}/2$ with the electroweak scale $\upsilon\simeq 246$ $\rm{GeV}$.

\subsection{WIMP DM relic density}

\begin{figure}[t]
\begin{centering}
\includegraphics[width=0.35\textwidth]{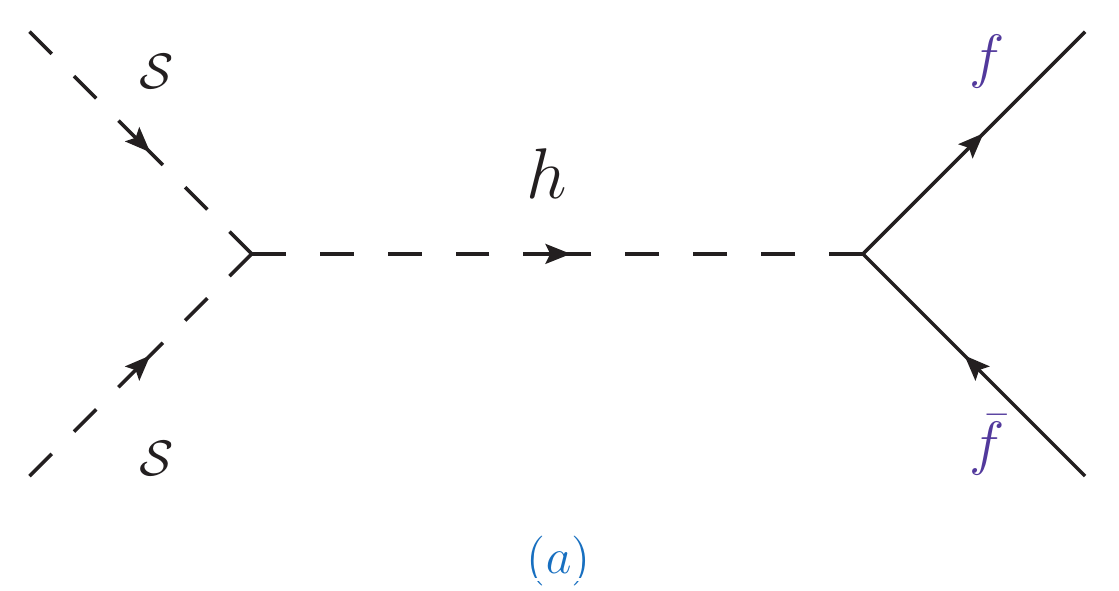}
\includegraphics[width=0.35\textwidth]{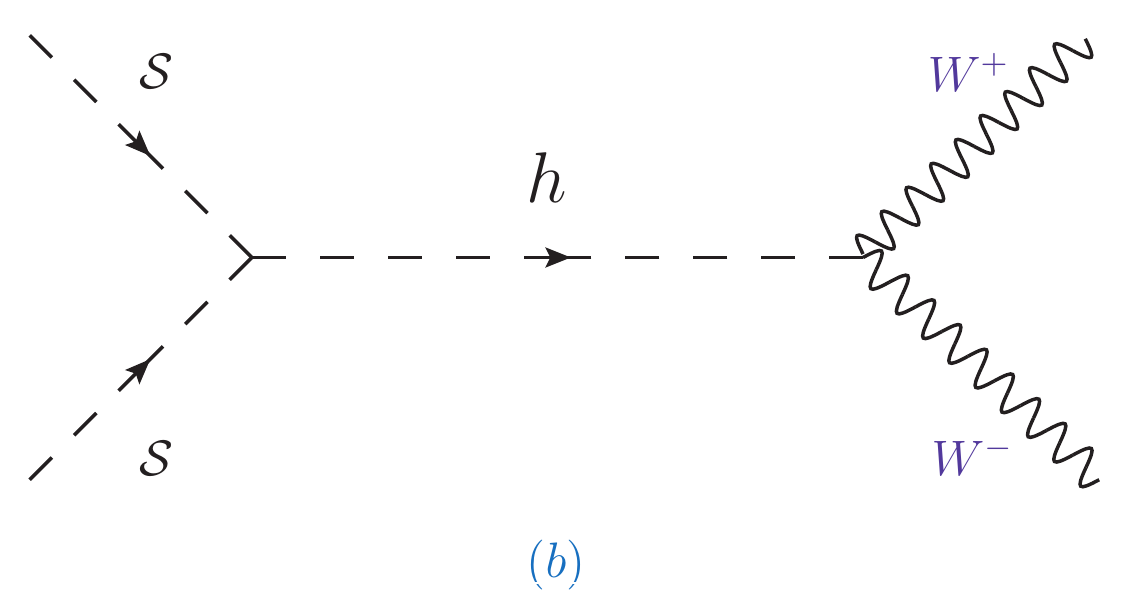}
\includegraphics[width=0.35\textwidth]{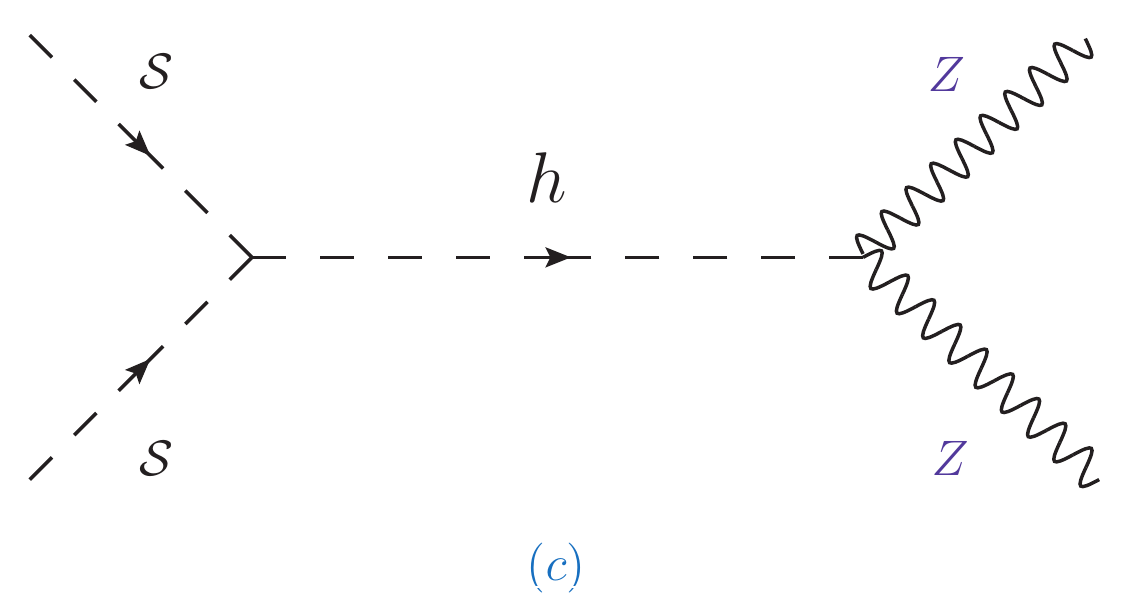}
\includegraphics[width=0.35\textwidth]{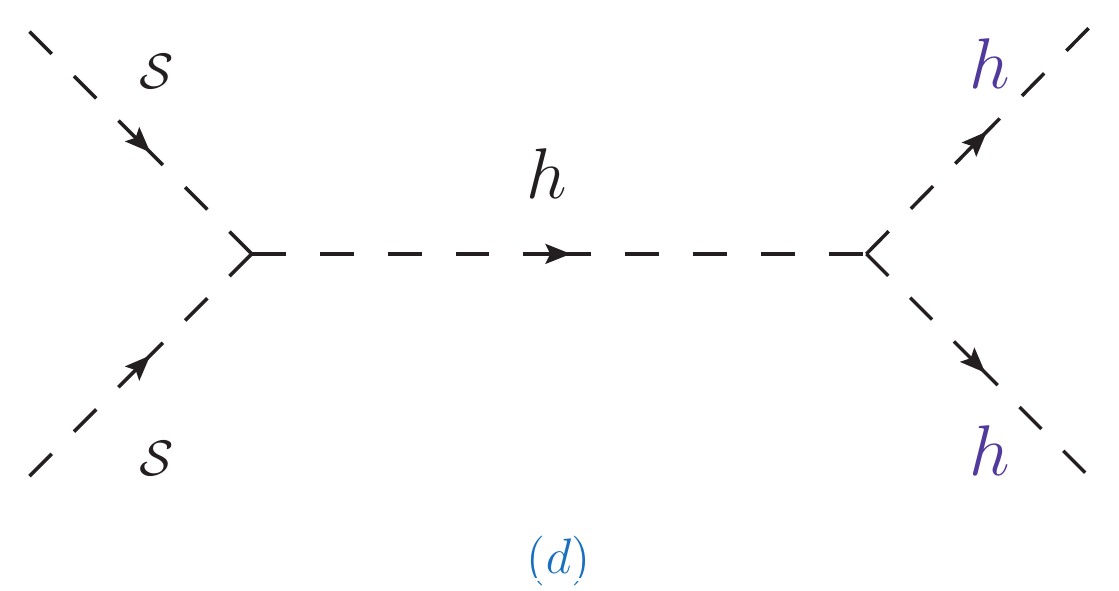}
\caption{Feynman diagrams for the scalar dark matter annihilation.}
\label{Fig:ssAni}
\end{centering}
\end{figure}

In this part, we will show the calculation technique in detail for the scalar DM thermal average annihilation cross section that is the important component of the scalar DM relic density, which can be written as~\cite{Gondolo:1990dk}:
\begin{eqnarray}{\label{EQ:sigmabar}}
\left\langle\sigma v\right\rangle=\frac{x}{8 m_{\mathcal{S}}^{5} K_{2}^{2}\left(x\right)}\int_{4 m_{\mathcal{S}}^{2}}^{\infty} (s-4 m_{\mathcal{S}}^{2}) \sqrt{s} K_{1}(x \sqrt{s} / m_{\mathcal{S}}) \sigma \mathrm{d} s.
\end{eqnarray}
Here $x=m_{\mathcal{S}} / T$ ($T$ is the temperature), $K_{1}(y)$ and $K_{2}(y)$ are the modified Bessel function of the second kind, and $s$ is the square of the center-of-mass energy. According to our scalar DM model Eq.~(\ref{V:HS}), there are four kinds of annihilation feynman diagrams which are shown in Fig.~\ref{Fig:ssAni}, then scalar dark matter annihilation cross section $\sigma$ can be obtained as,

\begin{eqnarray}{\label{sigmatot}}
\sigma =\sigma^{f\bar{f}}+\sigma ^{W^{+}W^{-}}+\sigma ^{ZZ}+\sigma^{hh}.
\end{eqnarray}
The cross section $\sigma^{XX}$ can be cast into the following form:
\begin{eqnarray}
\sigma^{XX} =\lambda_{h\mathcal{S}}^{2} \nu^{2}\left|D_{h}(s)\right|^{2} \Gamma_{h \rightarrow X X}(s),
\end{eqnarray}
where,
\begin{eqnarray}
\left|D_{h}(s)\right|^{2}  \equiv \frac{1}{\left(s-m_{h}^{2}\right)^{2}+m_{h}^{2} \Gamma_{h}^{2}}.
\end{eqnarray}
After utilizing the standard quantum field theory calculations, the cross section in right hands of Eq.~(\ref{sigmatot}) can be obtained. More explicitly, we give the final result as follows:
\begin{eqnarray}{\label{sigma}}
\sigma ^{f\bar{f}}&=&N_{C}^{f}\sum_{f} \frac{\lambda_{h\mathcal{S}}^{2} m_{f}^{2}\left(1-4 m_{f}^{2}/s\right)^{3/2}}{8 \pi \left(m_h^{2} \Gamma_{h}^{2}+\left(s-m_h^{2}\right)^{2}\right)},\\
\sigma ^{W^{+}W^{-}}&=&\frac{\lambda_{h\mathcal{S}}^{2} \left(s+12\frac{m_{W}^4}{s}-4m_{W}^{2}\right)\sqrt{1-4 m_{W}^{2}/s}}{16 \pi \left(m_h^{2} \Gamma_{h}^{2}+\left(s-m_h^{2}\right)^{2}\right)},\\
\sigma ^{ZZ}&=&\frac{\lambda_{h\mathcal{S}}^{2} \left(s+12\frac{m_{Z}^4}{s}-4m_{Z}^{2}\right)\sqrt{1-4 m_{Z}^{2}/s}}{32 \pi \left(m_h^{2} \Gamma_{h}^{2}+\left(s-m_h^{2}\right)^{2}\right)},\\
\sigma ^{hh}&=&\frac{9\lambda_{h\mathcal{S}}^{2} m_{h}^{4}\sqrt{1-4 m_{h}^{2}/s}}{32 \pi s\left(m_h^{2} \Gamma_{h}^{2}+\left(s-m_h^{2}\right)^{2}\right)}.
\end{eqnarray}

These analytic annihilation formulas indicate that $s$ must be more than twice the mass of a particle before the corresponding annihilation channel can be opened. For example, if $2m_W > s > 2m_{f\bar{f}}$, then only $\mathcal{S}\mathcal{S} \to f\bar{f}$ annihilation will be opened.

\begin{figure}[t]
\begin{centering}
\includegraphics[width=0.6\textwidth]{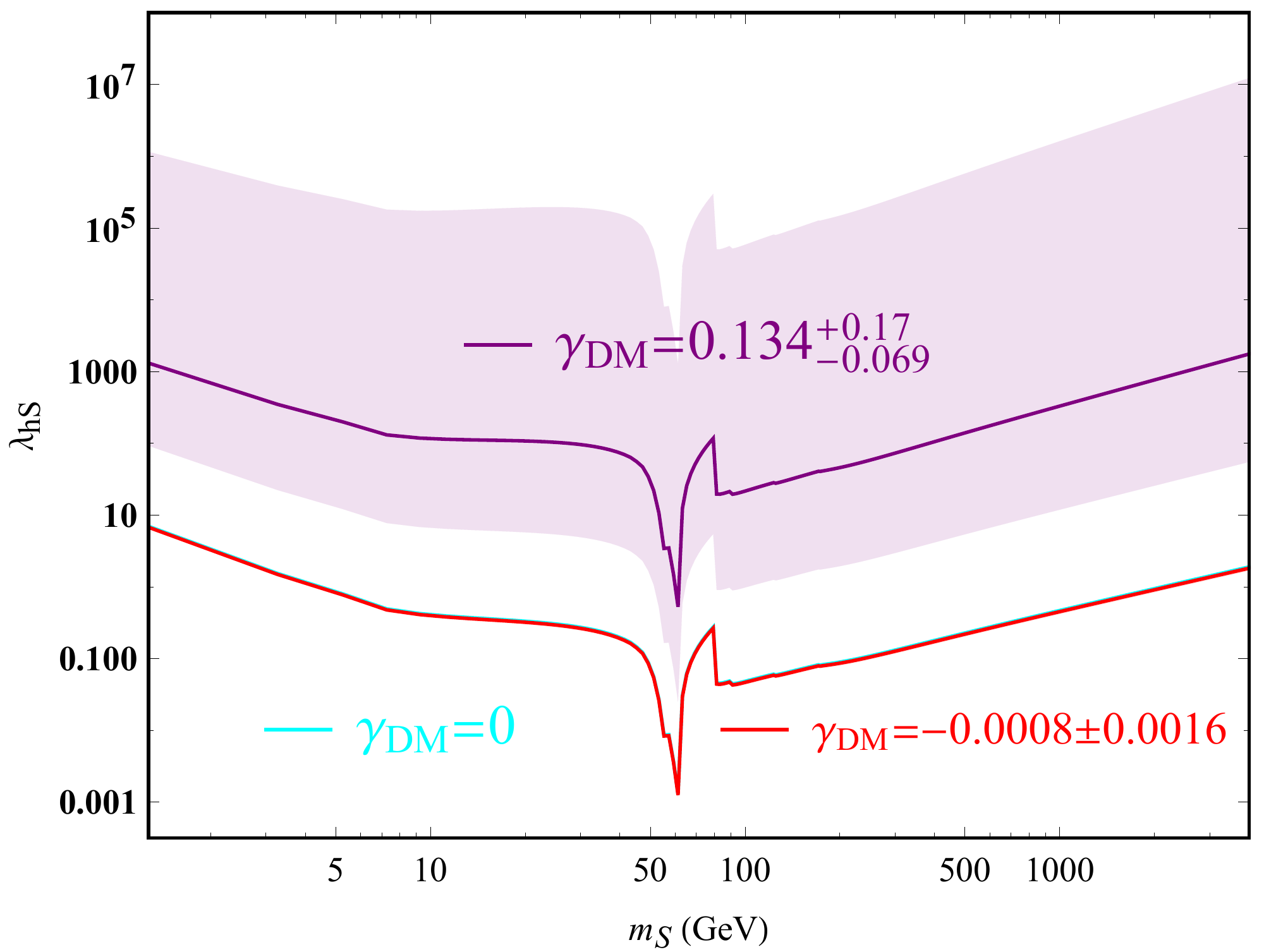}
\caption{Allowed parameter regions in the ($m_\mathcal{S}-\lambda_{h\mathcal{S}}$) plane under the constraint of dark matter relic density $\Omega{h^2}=0.1198 \pm 0.0012$~\cite{Aghanim:2018eyx}.}
\label{Fig:QlhsmsDR}
\end{centering}
\end{figure}

The newest Plank and WMAP data have measured the DM relic density in high precision~\cite{Aghanim:2018eyx}:
\begin{eqnarray}\label{EQ:DMrelden}
\Omega{h^2}=0.1198 \pm 0.0012.
\end{eqnarray}

For this scalar DM model, one can straightforward get the scalar DM relic density by inserting Eq.~(\ref{EQ:sigmabar}) into Eq.~(\ref{EQ:Q1DMRD}). We find that only the DM mass $m_{\mathcal{S}}$, the coupling strength $\gamma_{DM}$ and the coupling constant $\lambda_{h\mathcal{S}}$ are sensitive to the scalar DM relic density.

We show the allowed parameter regions in the ($m_{\mathcal{S}}-\lambda_{h\mathcal{S}}$) plane in Fig.~\ref{Fig:QlhsmsDR}, where the bounds come from the DM relic density Eq.~(\ref{EQ:DMrelden}). In Fig.~\ref{Fig:QlhsmsDR}, the cyan shad bands represent $\gamma_{DM}=0$ that means there is a vanishing interaction between DE and scalar DM. The purple shad bands represent $\gamma_{DM} = 0.134 ^{+ 0.17}_{-0.069}$ case which is obtained from the (SGL+SNe+Hz) cosmological observations bounds. In other words, there is an interaction between DE and scalar DM, and the energy of DE converts into that of the scalar DM. And $\lambda_{h\mathcal{S}}$ decreases compared to the $\gamma_{DM} = 0$ case when scalar DM relic density limit is taken into account. All the solid lines here represent the central values.

The red shad bands represent the case of $\gamma_{DM}=-0.0008\pm 0.0016$ that is obtained by the limitation of the cosmological observation amount (CMB+BAO+SNe). The negative coupling $\gamma_{DM}$ means not only there is an interaction between DE and DM but also the energy of DM flows to that of the DE. After taking into account the limitation of the relic density of DM, $\lambda_{h\mathcal{S}}$ will be increased compared with the vanishing interaction scenario. The coupling of Higgs and scalar DM $\lambda_{h\mathcal{S}}$ is too large to be reasonable and even is excluded completely, which can also be seen from Fig.~\ref{Fig:QlhsmsDR}.

It should be emphasized that although the $\gamma_{DM}$ obtained by the cosmological observation of (SGL+SNe+Hz) is large and beyond the range of $x_f$ set in the previous section, we found that in this case, $x_f$ still remains around $20$.

In addition, we can see that when all the lines have an obvious jump at $m_s\approx80~\rm{GeV}$, the reason is that the $\mathcal{S}\mathcal{S}\to W^+W^-$ channel annihilation will be opened when $s>2 m_W$. The contribution of this annihilation channel is larger and the discontinuity is very obvious. In fact, there will be a break in the lines as long as a channel is closed, which is not obvious due to the little difference in contribution between adjacent channels.

\subsection{DM direct detection}

The spin-independent DM-nucleon scattering cross section for the scalar DM model has the form of~\cite{He:2016mls}:
\begin{eqnarray}{\label{crosssection}}
\sigma_{\text{SI}}= \frac{\lambda_{h\mathcal{S}}^{2}f^{2}_{N}\mu^{2}m^{2}_{N}}{4\pi m^{4}_{h}m_{\mathcal{S}}^{2}},
\end{eqnarray}
where the DM-nucleon reduced mass $\mu=m_{N}m_{\mathcal{S}}/(m_{N}+m_{\mathcal{S}})$ with $m_{N}$ being the nucleon mass, the Higgs mass $m_{h}=125$ $\rm{GeV}$~\cite{Aad:2012tfa,Chatrchyan:2012ufa}, the hadron matrix element $f_{N}\simeq0.3$~\cite{Cline:2013gha}. $m_{\mathcal{S}}$ and $\lambda_{h\mathcal{S}}$ are the DM mass and the coupling constant respectively, and they depend on the constructed scalar DM model.

\begin{figure}[t]
\begin{centering}
\includegraphics[width=0.6\textwidth]{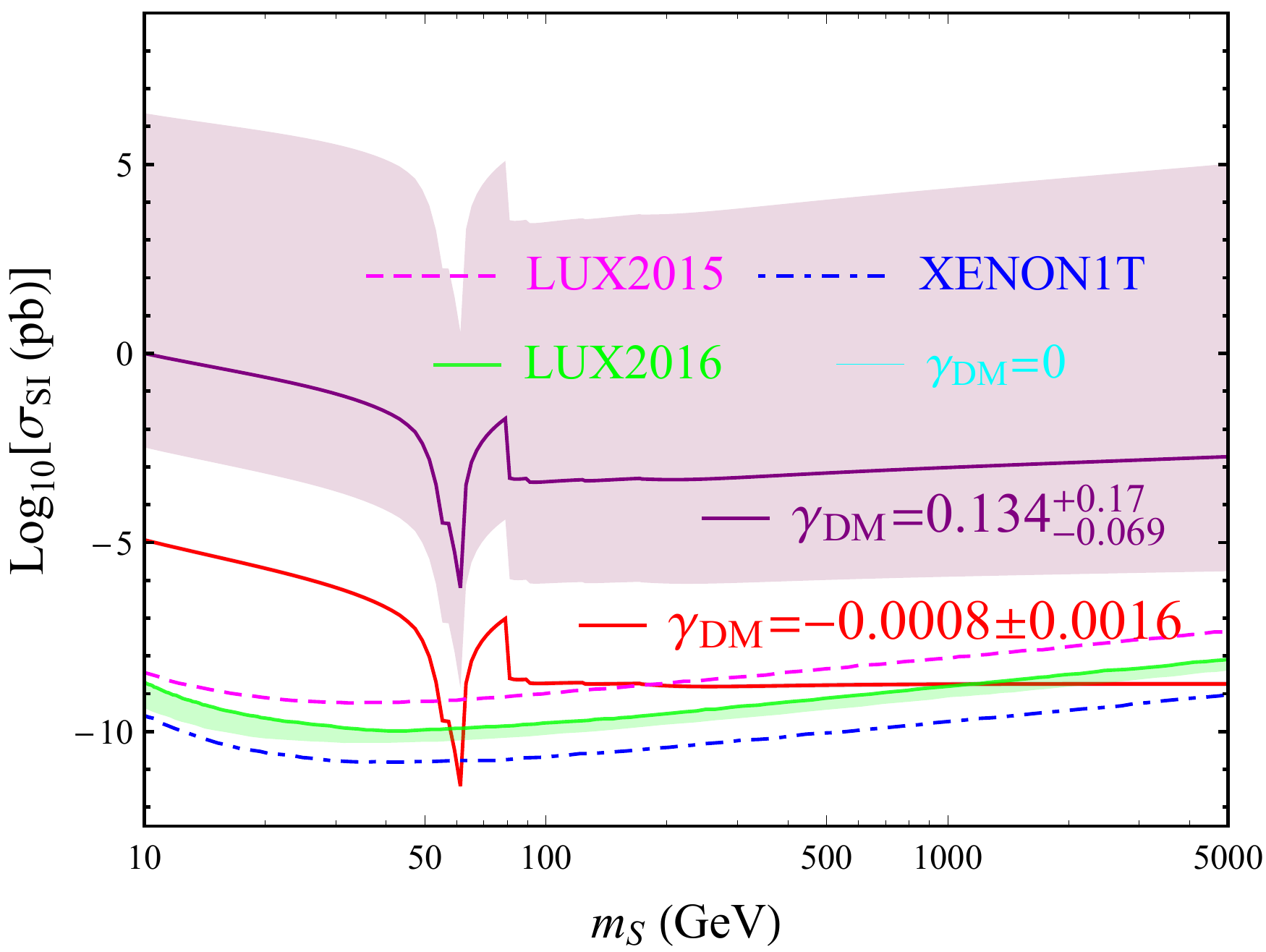}
\caption{DM-nucleon scattering cross section $\sigma_{SI}$ as the function of DM mass for several $\gamma_{DM}$. The magenta dashed line, purple line, and blue dot-dashed line represent the bounds from the LUX2015~\cite{Aprile:2015uzo}, LUX2016~\cite{Aprile:2015uzo} and Xenon1T~\cite{Aprile:2015uzo}, respectively. And the purple shad band refers to $2\sigma$ deviation to their central values that were chosen by authors in Ref.~\cite{He:2016mls,Escudero:2016gzx}.} \label{Fig:msSIQ1}
\end{centering}
\end{figure}

The spin-independent DM-nucleon mesh cross section as a function with $m_{\mathcal{S}}$ is shown in Fig.~\ref{Fig:msSIQ1}.
The limitations of the direct DM detection experiments include LUX2015, LUX2016, XENON1T,
which are marked by the magenta dashed, green shaded bands, and blue dot-dashed line, respectively.
While the cyan, red, and purple shaded bands are the $\gamma_{DM}=0$, $\gamma_{DM}=-0.0008\pm 0.0016$ and $\gamma_{DM}= 0.134 ^{+ 0.17}_{-0.069}$ cases, respectively. Note that the cyan shaded bands are completely covered by red shaded bands.

Compared with $\gamma_{DM}=0$, i.e., the DM and DE have vanishing interaction, Fig.~\ref{Fig:msSIQ1} indicates that:
\begin{itemize}
  \item When $\gamma_{DM} = 0.134 ^{+ 0.17}_{-0.069}$, the feasible parameter spaces of spin-independent DM-nucleon cross section shifts upward,
after considering both the direct detection experiments and the DM relic density bounds, the scalar DM model will be completely eliminated.
  \item When $\gamma_{DM}=-0.0008\pm0.0016$, the shape will be extended.
Also, after taking into account the direct detection experiment and DM relic density limitation, the feasible parameter space of the scalar DM model is increased.
\end{itemize}
For example, under the limits of LUX2016 experiment~\cite{Aprile:2015uzo}, the increase is not obvious in the resonant mass region, but it is relatively obvious in the large mass regions. More specifically, it can be obtained from Table.~\ref{VR}.

\begin{table}
\begin{center}
\begin{tabular}{ccc}
\hline
Viable regions                                            &~Resonant mass regions~                           &~High mass regions\\
\hline
$\gamma_{DM} = 0.134 ^{+ 0.17}_{-0.069}$   &        $-$                                           &  $-$              \\
$\gamma_{DM} =0$                          &$59~ \rm{GeV} \lesssim m_{\mathcal{S}} \lesssim 63~\rm{GeV}$        &  $m_{\mathcal{S}}\gtrsim 1.5~\rm{TeV}$  \\
$\gamma_{DM}=-0.0008\pm 0.0016$           &$58.5~ \rm{GeV} \lesssim m_{\mathcal{S}} \lesssim 63.5~\rm{GeV}$          &  $m_{\mathcal{S}} \gtrsim 1.45~\rm{TeV}$ \\
\hline
\end{tabular}
\caption{The viable regions in the resonant mass and high mass regions under the limits of LUX2016 for the three $\gamma_{DM}$ cases~\cite{Aprile:2015uzo}. }
\label{VR}
\end{center}
\end{table}

Therefore, if (SGL+SNe+Hz) and (CMB+BAO+SNe) cosmological observation data are completely credible within the errors, but they lead to a completely contradictory conclusion, then there may be something wrong with the scalar DM model.
On the contrary, if the scalar DM dark matter model is correct, then there must be something wrong for the (SGL+SNe+Hz) experimental data, while the correctness of the (CMB+BAO+SNe) cosmological observation data need to be further confirmed.

\subsection{DM indirect detection}

After discussing the influence of the added interaction term $Q_{DM}$ on the DM direct detection, in this section, we will turn to its impact on another kind of DM detection, i.e., DM indirect detection. Higgs invisible decay is an important part of the DM indirect detection.
According to the potential Eq.~\ref{Lag2}, the Higgs invisible decay channel is $h \to \mathcal{SS}$, and then the corresponding decay width is expressed as:
\begin{eqnarray}
\Gamma(h\to \mathcal{SS})=\frac{\lambda_{h\mathcal{S}}^2v^2}{32\pi m_h}\sqrt{1-\frac{4m_{\mathcal{S}}^2}{m_h^2}}
\end{eqnarray}
To make sense of this equation, in other words, to make Higgs invisible decay ($h \to \mathcal{SS}$) occur, thus $m_{\mathcal{S}}<1/2 m_h$.

\begin{figure}[t]
\begin{centering}
\includegraphics[width=0.6\textwidth]{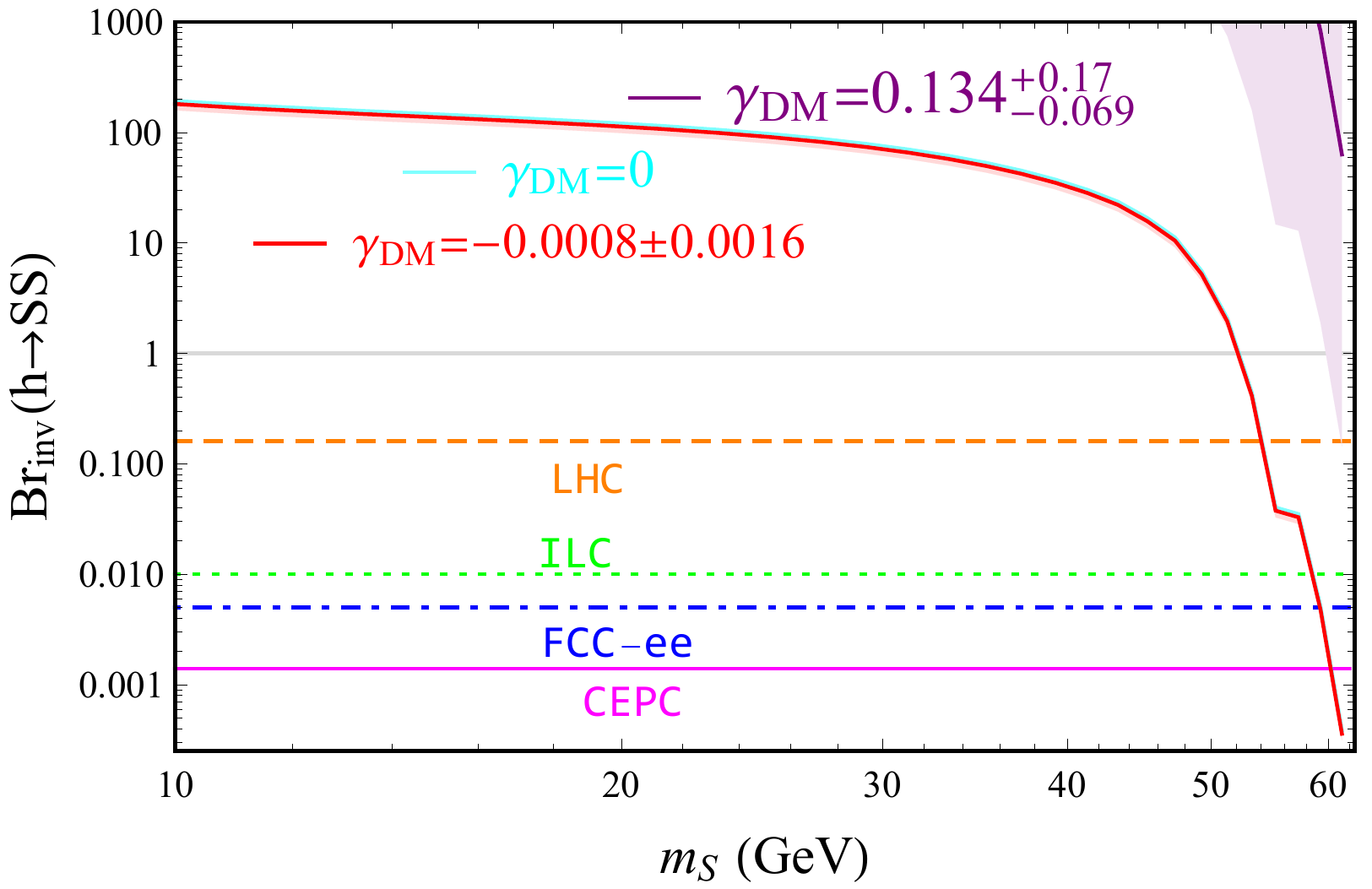}
\caption{The branching ratio of the Higgs invisible decay.}
\label{Fig:msBr}
\end{centering}
\end{figure}

The interaction term $Q_{DM}$ will make an effect on the interaction coupling $\lambda_{h\mathcal{S}}$, as can be seen in Fig.~\ref{Fig:QlhsmsDR}.
This effect can be further transmitted to the Higgs invisible decay width ($\Gamma(h\to \mathcal{SS})$) through the coupling $\lambda_{h\mathcal{S}}$.
Applying the total Higgs decay width ($\Gamma_h\simeq4.15~\rm{MeV}$), we plot the branching ratio of the Higgs invisible decay in Fig.~\ref{Fig:msBr}.
In which, we also show the limits from the LHC $Br_{inv}<16\%$~\cite{Khachatryan:2016vau}, ILC $Br_{inv}<1\%$~\cite{Fujii:2015jha}, FCC-ee $Br_{inv}<0.5\%$~\cite{Gomez-Ceballos:2013zzn,dEnterria:2016sca}, and CEPC $Br_{inv}<0.14\%$~\cite{CEPC-SPPCStudyGroup:2015csa}.
The case ($\gamma_{DM} = 0.134 ^{+ 0.17}_{-0.069}$) is almost excluded by all the experiments, while for the remaining two cases, the experimental limits ILC, FCC-ee, and CEPC lead to $58.5~\rm{GeV} \lesssim m_{\mathcal{S}}\lesssim62.5~\rm{GeV}$. Those are almost consistent with DM direct experiment (see Tab.~\ref{VR}).

\section{Summary}\label{sec:VI}

In this paper, motivated by the simplest interaction model between
DE and DM with interaction term $Q=3\gamma_{DM}H\rho_{DM}$, we
construct the corresponding interaction model of DE and WIMP DM, in
which the WIMP DM annihilation term and the interaction term between
DM and DE are considered simultaneously. We then discuss the
freeze-out parameter $x_f=m_{DM}/T$ with $T$ being the DM freeze-out
temperature and deduced the WIMP DM relic density in this new
scenario. The resultant DM relic density will be magnified by
$\frac{2-3\gamma_{DM}}{2}[{2\pi g_* m_{DM}^3}/{(45 s_0
x_f^3})]^{\gamma_{DM}}$ times. The new DM relic density opens a new
way to test WIMP DM models through the interaction strength
$\gamma_{DM}$, using the observational constraints from the latest
cosmological data.

For the interaction strength $\gamma_{DM}$, we employ cosmological
observations to constrain the interaction strength $\gamma_{DM}$ and
find that different observations may make a significant difference
for the interaction strength $\gamma_{DM}$. Specifically,
(SGL+SNe+Hz) and (CMB+BAO+SNe) cosmological observation data will
give out $\gamma_{DM}=0.134^{+0.17}_{-0.069}$ and
$\gamma_{DM}=-0.0008\pm0.0016$ respectively. When the interaction
strength is $\gamma_{DM}=0$, the interaction will vanish, and the
model will reduce to the standard $\Lambda$CDM model.

As an example, we analyze the case where WIMP DM is a scalar DM.
After further considering the constraints from the DM direct
detection experiment, DM indirect detection experiments and DM relic
density, we observe that the allowed parameter space of the scalar
DM model will be completely excluded for (SGL+SNe+Hz) cosmological
observation data, while it will be increased for (CMB+BAO+SNe)
cosmological observation data. Those two cosmological observation
data lead to a paradoxical conclusion. Thus, more accurate
predictions for $\gamma_{DM}$ based on the cosmological observations
data will provide us with a possible way to screen the WIMP DM
models.

\hspace{2cm}

\noindent{\bf Acknowledgement}: We are grateful to Prof. Yu-Feng
Zhou, Shuo Cao and Xiaolei Li for helpful communications and
discussions. This work was supported in part by Graduate Research
and Innovation Foundation of Chongqing, China (Grant No. CYS20272),
the China Postdoctoral Science Foundation under Grant No.
(2019TQ0329, 2020M670476), and the National Natural Science
Foundation of China under Grant No.11947302. Jia-Wei Zhang was
supported by the Natural Science Foundation of Chongqing under
Grants No.(cstc2018jcyjAX0713), the Science and Technology Research
Program of Chongqing Municipal Education Commission under Grant No.
KJQN202001541, and the Research Foundation of Chongqing University
of Science and Technology under Grant No. CK2016Z03.

\bibliographystyle{arxivref}

\end{document}